\documentclass[10pt]{iopart}
\usepackage{iopams}
\usepackage{graphicx}  
\usepackage{cite}
\usepackage[pdfstartview=FitH, pdfpagemode=None, pdfpagelayout=OneColumn, colorlinks=true, linkcolor=blue, citecolor = blue]{hyperref}
\newcommand{\text}[1]{\mathrm{#1}}
\begin{document}

\title{Ion dynamics driven by a strongly nonlinear plasma wake}

\author{V K Khudiakov$^{1,2}$, K V Lotov$^{1,2}$, M C Downer$^{3}$}
\address{$^1$Budker Institute of Nuclear Physics, Novosibirsk, Russia}
\address{$^2$Novosibirsk State University, Novosibirsk, Russia}
\address{$^3$University of Texas at Austin, Austin, TX, USA}
\vspace{10pt}
\begin{indented}
\item[] \today
\end{indented}

\begin{abstract}
In plasma wakefield accelerators, the wave excited in the plasma eventually breaks and leaves behind slowly changing fields and currents that perturb the ion density background. We study this process numerically using the example of a FACET experiment where the wave is excited by an electron bunch in the bubble regime in a radially bounded plasma. Four physical effects underlie the dynamics of ions: (1) attraction of ions toward the axis by the fields of the driver and the wave, resulting in formation of a density peak, (2) generation of ion-acoustic solitons following the decay of the density peak, (3) positive plasma charging after wave breaking, leading to acceleration of some ions in the radial direction, and (4) plasma pinching by the current generated during the wave-breaking. Interplay of these effects result in formation of various radial density profiles, which are difficult to produce in any other way.
\end{abstract}

%
\vspace{2pc}
\ioptwocol

\section{Introduction}

Acceleration of electrons and positrons in near-light-speed plasma waves is a rapidly developing field of research \cite{RMP81-1229,RAST9-63,RAST9-85,RMP90-035002,NJP23-031101}. The highest performing plasma accelerators are based on strongly nonlinear wakefield by a relativistic electron bunch or a high-intensity laser pulse in the blowout regime \cite{PRA44-6189,APB74-355}. In this case, the field of the driver is strong enough to completely expel electrons from its immediate wake, forming a so-called `bubble'.

Usually, in plasma-based accelerators, the accelerated beam (called a witness) follows closely behind the driver. The time between the two beams is too short for plasma ions to respond, except in the case of very dense electron drivers \cite{PRL95-195002,PRL104-155001,PRST-AB14-021303,PRL118-244801,PRAB20-111301} or special regimes in which a small perturbation of the ion density stabilizes the electron driver \cite{PRL121-264802}. For this reason, the ion dynamics has not attracted as much attention as other wakefield features, and has been studied mainly in the context of the wakefield lifetime \cite{PRL86-3332,PoP10-1124,PRL109-145005,PoP21-056705,PoP25-103103} or the overall energy balance in the system \cite{NatComm11-4753}. However, the ion motion can result in formation of exotic, slowly changing plasma density profiles, which may be useful for diagnostics \cite{PRX9-011046,NatComm11-4753}, driver guiding \cite{RMP81-1229}, beam stabilization \cite{PoP20-080701,PRL121-264801}, wakefield enhancement \cite{PPCF61-114003,PRL113-245003}, radiation generation \cite{PoP24-113110}, or making field structure favorable for positron acceleration \cite{PoP14-023101,PRAB22-081301,SRep4-4171,PPCF61-025012}. Ion density valleys or peaks can appear both in strongly nonlinear \cite{PRL105-195002,PRAB20-081004} and almost linear  \cite{PRL86-3332,PoP10-1124,PRX9-011046,PoP25-103103,PPR28-125, PRL109-145005,PoP21-056705,NJP9-402} regimes and for all kinds of drivers.
As more dense drivers become available \cite{PRAB22-101301}, the importance of ion dynamics will increase.
Despite the visual similarity of the observed density structures, the mechanisms of their formation are different in each case. In this paper, we study the ion dynamics after the wakefield excitation by a dense electron beam in a radially bounded plasma in the bubble regime. In this regime, the driver can act on the ions not only by means of the excited wave, but also directly by its own field and by the field of the current arising in the plasma. %

The paper is organized as follows. In section~\ref{intro2}, we describe the problem under study and the methods. Then we focus our attention on four effects, some of which are specific to electron drivers in radially bounded plasmas. The first effect is that ions from some paraxial region receive a negative radial momentum and form a density peak on the axis (section~\ref{sec2}). In our case, both the electric field of the beam and the ponderomotive force of the plasma wave are responsible for the peak formation. The second effect is the appearance of ion-acoustic solitons (section~\ref{sec3}). The on-axis density peak eventually breaks up into several smaller, diverging peaks, which propagate at the ion-sound velocity. The third effect is related to fast electrons that appear when the bubble collapses (section~\ref{sec4}). These electrons escape from the plasma column and leave an uncompensated ion charge behind, which, in turn, pushes outer ion layers radially. The escape of fast electrons also creates a compensating current in the plasma, which ultimately leads to the fourth effect (section~\ref{sec5}): the bulk of the ions contracts to the axis, forming a high-amplitude compression wave. In section~\ref{sec6}, we summarize the main findings.

\begin{figure}[tb]
\centering\includegraphics{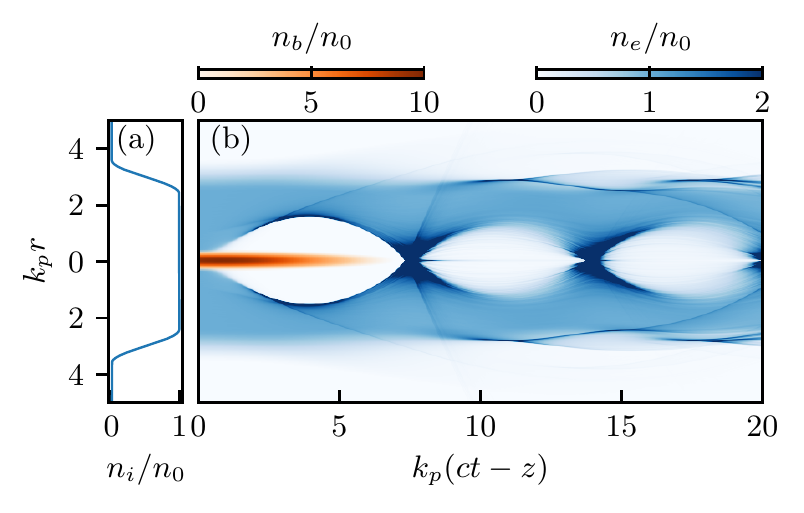}
\caption{ Wakefield excitation by the equilibrium beam: (a) radial dependence of the plasma ion density $n_i$ shortly after ionization and (b) the simulated density of the electron bunch $n_b$ (orange) and plasma electrons $n_e$ (blue) during the first three wakefield periods. The bunch propagates to the left.}\label{fig1-scheme}
\end{figure}

\section{Statement of the problem}\label{intro2}

We consider the case that corresponds to the recent E224 experiment \cite{NatComm11-4753} at the SLAC Facility for Advanced aCcelerator Experimental Tests (FACET) \cite{NJP12-055030}. In the E224 experiment, the temporal evolution of the plasma density profile was measured and compared to numerical simulations. The achieved quantitative agreement proved the validity of the simulation method, so we use the same approach and the same set of parameters in our study.

\begin{figure*}[tb]
	\centering
	\includegraphics[width = \textwidth]{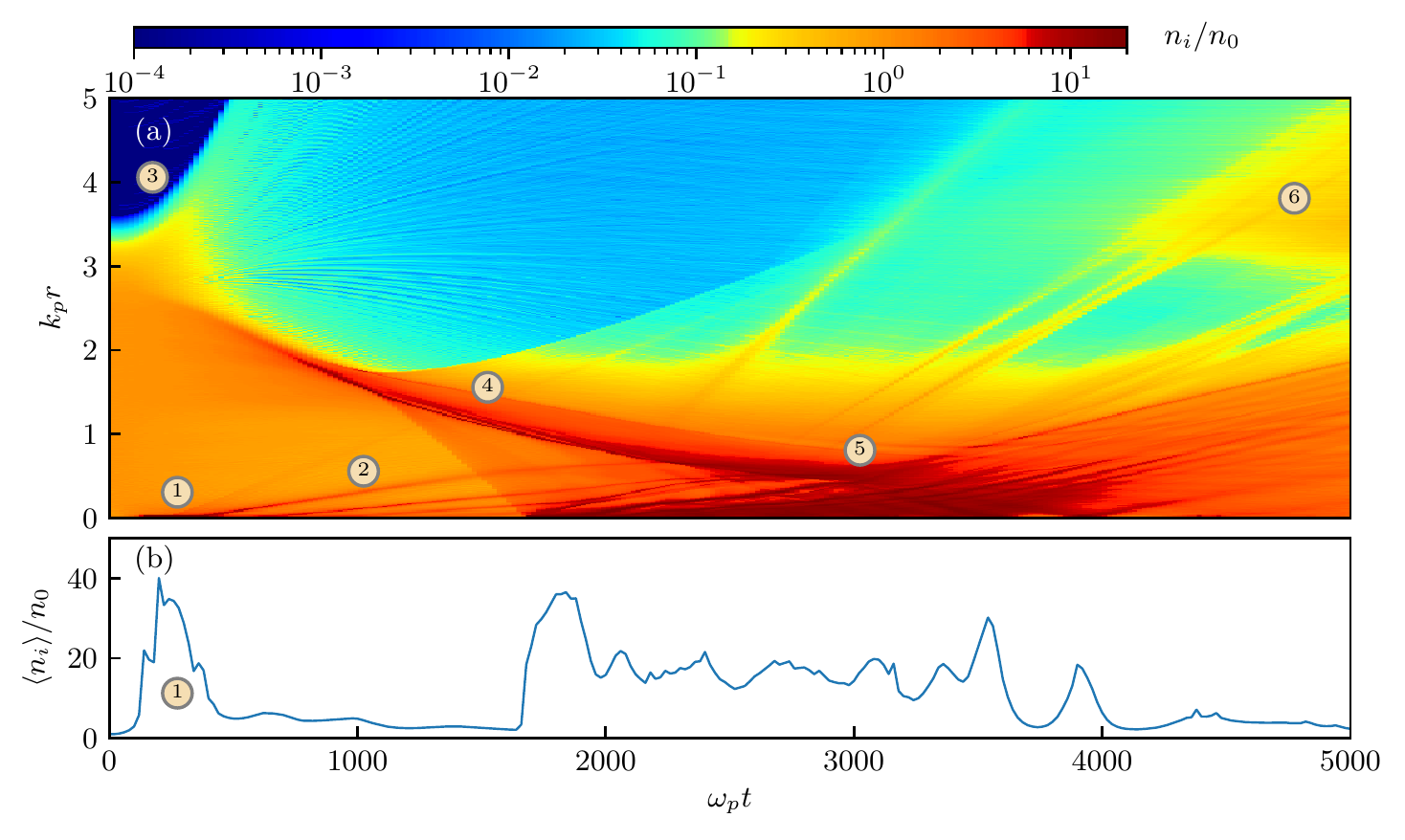}
	\caption{(a) The ion density map $n_i (r, t)$. The numbers in circles indicate the features discussed in the text and detailed in Supplementary Materials, movie~1. (b) The average ion density $\langle {n}_i \rangle (t)$ in the region $r < 0.05 k_p^{-1}$. The averaged density does not depend on the simulation grid step, unlike the density on the axis.}\label{fig2-big_dens}
\end{figure*}

An electron bunch of energy 20\,GeV, charge 2\,nC, root-mean-squared (rms) radius $30\,\mu$m and rms length $55\,\mu$m  passes through a chamber filled with lithium vapor of atomic density $n_0=8\times 10^{16}\,\rm{cm}^{-3}$ \cite{IEEE-PS27-791,Nat.515-92}. The bunch head field-ionizes the lithium and creates a plasma column of radius about $40\,\mu$m (figure \ref{fig1-scheme}). The plasma focuses the rest of the bunch down to the equilibrium radius of about $4\,\mu$m. After a short period of initial equilibration, most of the bunch propagates almost without changing its shape and excites a strongly nonlinear plasma wave in a plasma of constant radius. The equilibration stage (first 32\,cm of bunch propagation in the plasma) is simulated with a 2d3v (axisymmetric) version of the particle-in-cell code OSIRIS \cite{NatComm11-4753,PPCF55-124011}. Then, the equilibrium profiles of plasma and beam are imported into the quasistatic axisymmetric 2d3v code LCODE \cite{PRST-AB6-061301,NIMA-829-350} as initial conditions, and the simulations continue up to thousands of wakefield periods at a constant longitudinal position $z$. In the equilibrium, approximately 60\% of the beam (1.2 nC) propagates in the plasma and participates in driving the plasma wave.

To present the results in a more general form, we measure times in units of $\omega_p^{-1}$, distances in $k_p^{-1} = c/\omega_p$, fields in $E_0 = m_ec\omega_p/e$, and densities in $n_0$, where $\omega_p = \sqrt{4\pi n_0e^2/m_e}$ is the plasma frequency, $m_e$ is the electron mass, $e$ is the elementary charge, and $c$ is the speed of light. For $n_0=8\times 10^{16}\,\rm{cm}^{-3}$, $\omega_p^{-1} \approx 60$\,fs, $k_p^{-1} \approx 19\,\mu$m, and $E_0 \approx 27\,\rm{GV/m} \approx 0.9$\,MGs. For lithium ions of the mass $m_i \approx 12850\,m_e$, the characteristic timescale of ion response is $\tau_i = \omega_p^{-1} \sqrt{m_i/m_e} \approx 113\,\omega_p^{-1} \approx 7$\,ps.

The radius of the simulation window is $500\,k_p^{-1}$, the radial grid step is $0.005\,k_p^{-1}$, and the time step is $0.005\,\omega_p^{-1}$. The plasma consists of two species (electrons and ions) with $2.5\times 10^5$ equally weighted macro-particles in each. The number of plasma macro-particles slightly increases due to impact ionization of the surrounding neutral vapor \cite{NatComm11-4753}, but this effect is not significant at the times considered here.

\section{On-axis density peak}\label{sec2}

Formation of the ion density peak near the axis (figure~\ref{fig2-big_dens}, feature 1) is routinely observed in simulations of various plasma-based accelerator schemes \cite{PRL86-3332,PPR28-125,PoP10-1124,PRL109-145005,NJP16-033031,PRAB20-081004,PoP25-103103,PRX9-011046}. In the bubble regime, the peak appears because of the strong radial electric field of an electron beam located inside the bubble \cite{PRL105-195002,PRL95-195002}. In turn, in moderately nonlinear wakes, the inward radial force on the ions arises from specific properties of the plasma wave. The ponderomotive force of the wave is such that it pushes the plasma towards the axis \cite{PRL86-3332,PoP10-1124}. Another explanation of the same effect is that plasma electrons have different oscillation amplitudes at different radii, which leads to an average charge separation and, consequently, to a nonzero radial field experienced by the ions \cite{PRL109-145005,PoP21-056705}. The ions gain the radial momentum gradually over many wave periods rather than during a short time of driver passage. 

\begin{figure}[tb]
	\centering
	\includegraphics{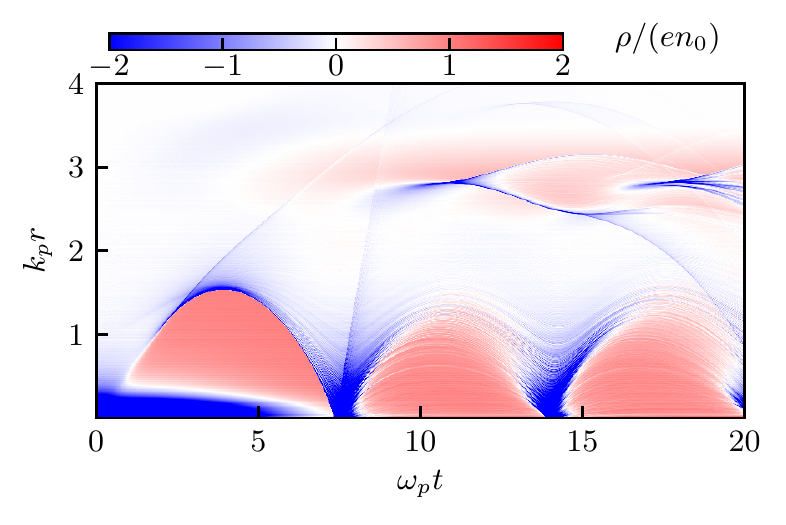}
	\caption{The total charge density of plasma and bunch, $\rho = e(n_i - n_e - n_b)$.}\label{fig3-total}
\end{figure}

\begin{figure}[tb]
	\centering
	\includegraphics[width=\linewidth]{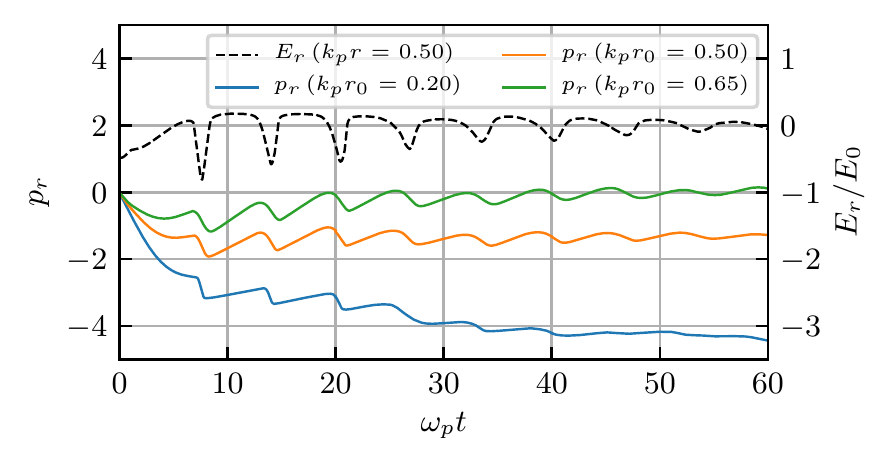}
	\caption{The radial momentum $p_r (t)$ of ions initially located at different radial positions $r_0$ (colored lines) and the radial electric field $E_r (t)$ at $r=0.5\,k_p^{-1}$ (black dashed line).}\label{fig4-pr-t}
\end{figure}

\begin{figure}[tb]
	\centering
	\includegraphics{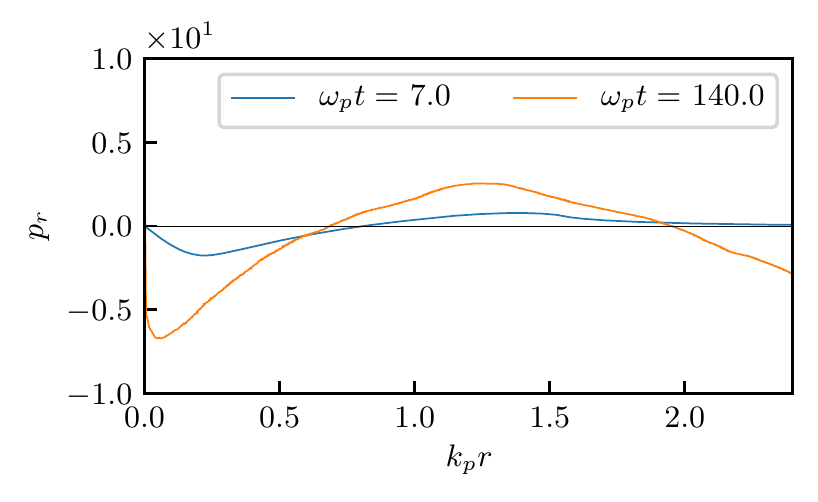}
	\caption{The radial ion momentum $p_r (r)$ at different times. The line $\omega_p t = 7$ shows the momentum gained mainly because of the beam field; the line $\omega_p t = 140$ corresponds to the arrival of near-axis ions to the axis and shows the combined effect of the beam and the wave.}\label{fig5-pr-r}
\end{figure}

In the considered case of a dense electron driver, the contributions of both the driver and the wave are important. The total charge density in the axial region is such that there is a large negative charge of the beam in the first bubble, and then narrow negatively charged and extended positively charged regions alternate (figure~\ref{fig3-total}). The ions first receive an inward push from the driver and then experience an oscillating force with a non-zero average (figure~\ref{fig4-pr-t}). Relative contributions of the initial push and wave force are comparable in value. In the considered case, the wave contribution is roughly 5 times stronger: the line $\omega_p t = 7$ in figure~\ref{fig5-pr-r} characterizes the initial push, while the difference between this line and the line $\omega_p t = 140$ is due to the wave force. As a consequence, the time when ions from the near-axis region reach the axis is shorter than it would be if they received only the initial push, $140\,\omega_p^{-1}$ instead of $10^3\omega_p^{-1}$. The latter number follows from the slope of the blue line in figure~\ref{fig5-pr-r}.

\begin{figure}[tb]
	\centering
	\includegraphics{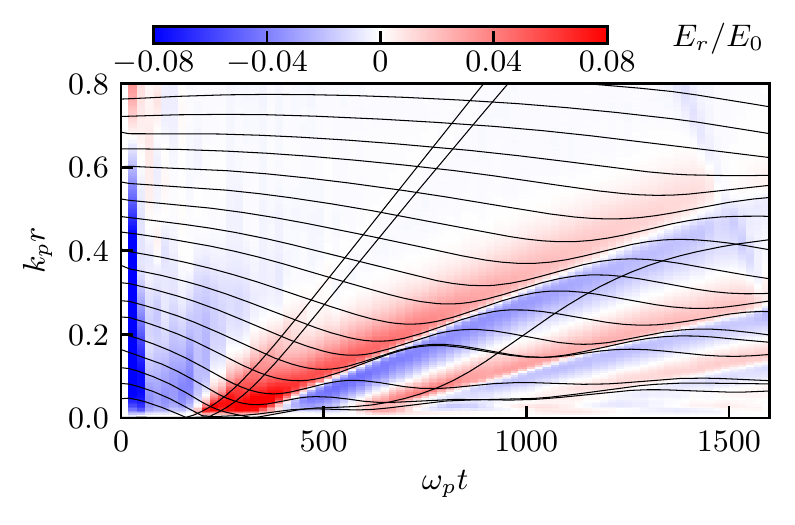}
	\caption{Equidistantly selected ion trajectories $r(t)$ plotted over the map of the radial electric field $E_r (r,t)$. The field is time-averaged over $20\omega_p^{-1}$ to reduce the effect of plasma oscillations.}\label{fig6-ion_traj}
\end{figure}

\section{Ion-acoustic solitons}\label{sec3}

\begin{figure}[tb]
	\centering
	\includegraphics[width=\linewidth]{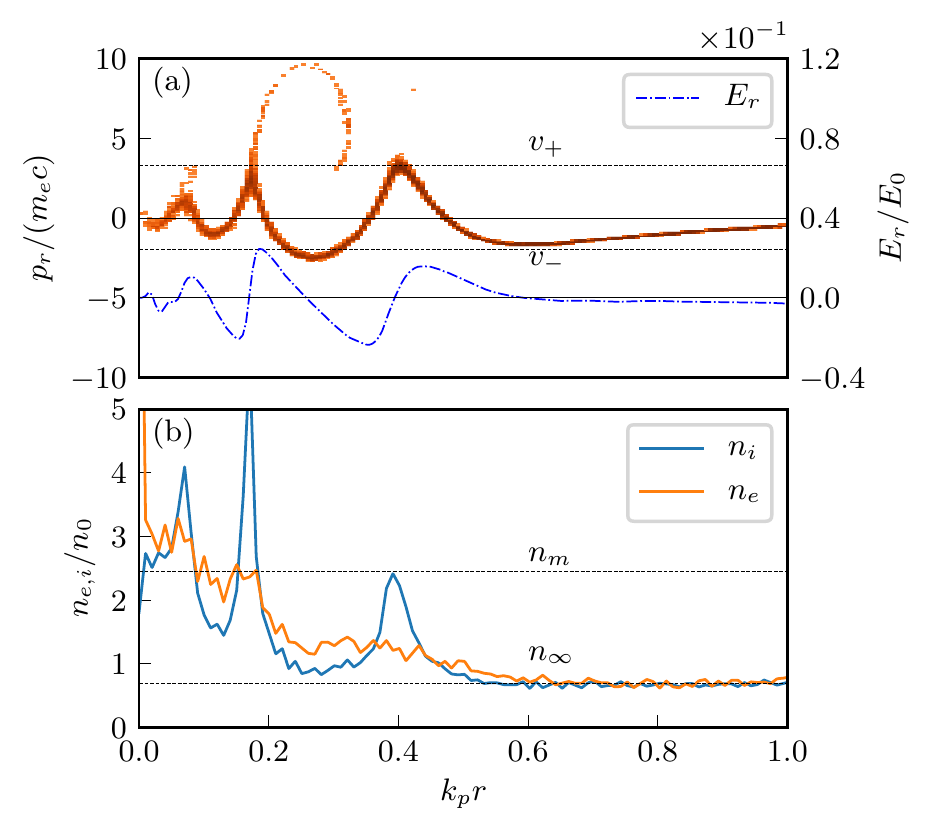}
	\caption{(a) The phase portrait of plasma ions (orange points) and the radial electric field $E_r$ (blue dashed line) at $\omega_p t = 1200$. The electric field is time-averaged over $20\omega_p^{-1}$ to reduce the effect of plasma oscillations. (b) The corresponding radial dependencies of electron ($n_e$) and ion ($n_i$) densities. The electron density is smoother because of the high electron temperature, and this leads to an ambipolar potential. The dashed lines show the soliton parameters discussed in the text.}\label{fig7-sol-amp}
\end{figure}

The movement of the ions toward the axis, followed by the formation of a density peak there, gives rise to ion-acoustic solitons, which are seen as diverging density ridges in figure~\ref{fig2-big_dens} (feature~2). When the first ions cross the axis, they not only create the density peak, but also a positive ambipolar potential, which attracts additional electrons to maintain average plasma quasi-neutrality. The resulting radial electric field reverses the direction of the next portion of ions (figure~\ref{fig6-ion_traj}). The counterstreaming ions form an off-axis density peak accompanied by an ambipolar potential and an electric field directed away from the peak. The peak moves radially outward against the background of inward-moving ions (figure~\ref{fig7-sol-amp}). As the soliton moves away from the axis, there is nothing to prevent the next portion of ions from passing through the soliton, reaching the axis and forming the next density peak there, which gives rise to the next soliton, and so on. Soliton generation continues as long at there is ion motion towards the axis. The dynamics of ion and electron density, ion velocity and radial electric field during soliton formation is detailed in the Supplementary materials, movie~1. Similar solitons, but not the mechanism of their formation, were also observed in simulations \cite{PRAB20-081004}.

To make sure that the observed feature really behaves as a soliton, we compare its velocity with the velocity expected for a soliton. We follow the approach outlined in Ref.\,\cite{Solitons} and start from the hydrodynamic equations for one-dimensional ion motion:
\begin{eqnarray}
\frac{\partial n}{\partial t} & = & -\frac{\partial (nv)}{\partial x}, \\
\frac{\partial v}{\partial t} & = & -\frac{\partial}{\partial x}\left(\frac{v^2}{2} + \frac{\phi}{\tilde{m}_i}\right), \\
\frac{\partial^2 \phi}{\partial x^2} & = & n_\infty e^{\phi/T} - n,
\end{eqnarray}
where we use dimensionless quantities: ion density $n = n_i/n_0$, electrostatic potential $\phi$ (normalized to $m_ec^2/e$), ion velocity $v = v_i/c$, unperturbed ion density $n_\infty$, electron temperature $T = T_e/(m_ec^2)$, and ion mass $\tilde{m}_i = m_i/m_e$. The ions are cold.

We are looking for a stationary solution, vanishing at infinity, propagating with velocity $u$ and depending on $\zeta = x - ut$. Then the equations can be integrated:
\begin{eqnarray}
n = \frac{n_\infty}{\sqrt{1 - \phi/\phi_0}} = \frac{n_\infty u}{u - v}, \\ \label{eq:nv}
v = u \left( 1 - \sqrt{1 - \phi/\phi_0} \right) = \frac{n - n_\infty}{n}u,\\
\left( \frac{\partial \phi}{\partial \zeta} \right)^2 = 2 n_\infty\left(T \bigl( e^{\phi/T} - 1 \bigr) + 
2\phi_0\bigl( \sqrt{1-\phi/\phi_0}- 1 \bigr) \right),
\end{eqnarray}
where $\phi_0 = \tilde{m}_i u^2/2$.

Let $n_m$, $v_m$ and $\phi_m$ be the maximum values of the solution reached where $\partial \phi / \partial \zeta = 0$. They are related as follows:
\begin{eqnarray}
\phi_m = \phi_0\frac{n_m^2 - n_\infty^2}{n_m^2},\\ \label{eq:T}
T \left( e^{\phi_m/T} - 1 \right) + 2\phi_0 \left( \sqrt{1-\phi_m/\phi_0} - 1 \right) = 0.
\end{eqnarray}

We analyze the outermost soliton, as its properties are least affected by the cylindricity. Its numerical parameters can be taken from figure \ref{fig7-sol-amp}: background ion velocity $v_{-} = -1.6\times 10^{-4}$, maximum ion velocity $v_{+} = 2.6\times 10^{-4}$, background density $n_\infty = 0.7$, and peak ion density $n_m = 2.45$. From the equation (\ref{eq:nv}), we find the soliton propagation velocity with respect to the ion background
\begin{equation}\label{eq:u}
u = \frac{n_m}{n_m - n_\infty}(v_+ - v_-)= 6.0\times 10^{-4}
\end{equation}
and in the laboratory frame
\begin{equation}\label{eq:u_l}
u_l = u + v_- = 4.2\times 10^{-4}.
\end{equation}
On the other hand, the propagation velocity of the density peak can be directly measured in figure \ref{fig2-big_dens}(a):
\begin{equation}
u_{l,\:sim} = 4.4\times 10^{-4},
\end{equation}
which agrees with the theoretical value (\ref{eq:u_l}).

\begin{figure}[tb]
	\centering
	\includegraphics{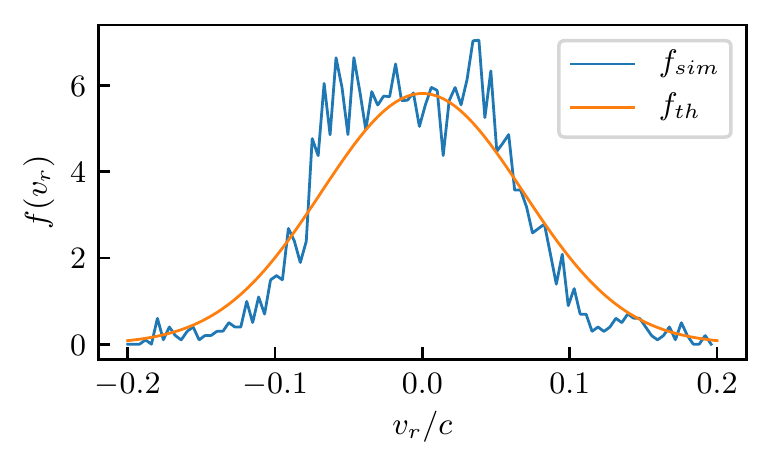}
	\caption{The electron distribution function $f_{sim}(v_r)$ observed in simulations near the outermost soliton at $\omega_p t = 1200$ and the Maxwell distribution $f_{th}$ for the estimated temperature $T_e = 2.4$\,keV.}\label{fig8-distr}
\end{figure}

The velocity and peak density of the soliton determine the maximum potential and electron temperature from the equation (\ref{eq:T}): $\phi_m = 3.9\times 10^{-3}$, $T = 4.7\times 10^{-3}$, $T_e = 2.4$\,keV. The ion sound velocity for this temperature is close to the soliton velocity (\ref{eq:u}):
\begin{equation}
    c_s = \sqrt{\frac{T_e}{m_i}} = 6.1\times 10^{-4} c.
\end{equation}
The Maxwell distribution for this temperature is close to the electron velocity distribution observed in simulations (figure \ref{fig8-distr}). Taken together, this proves that the observed features are ion-acoustic solitons. 

\begin{figure*}[tb]
	\centering
	\includegraphics[width = \textwidth]{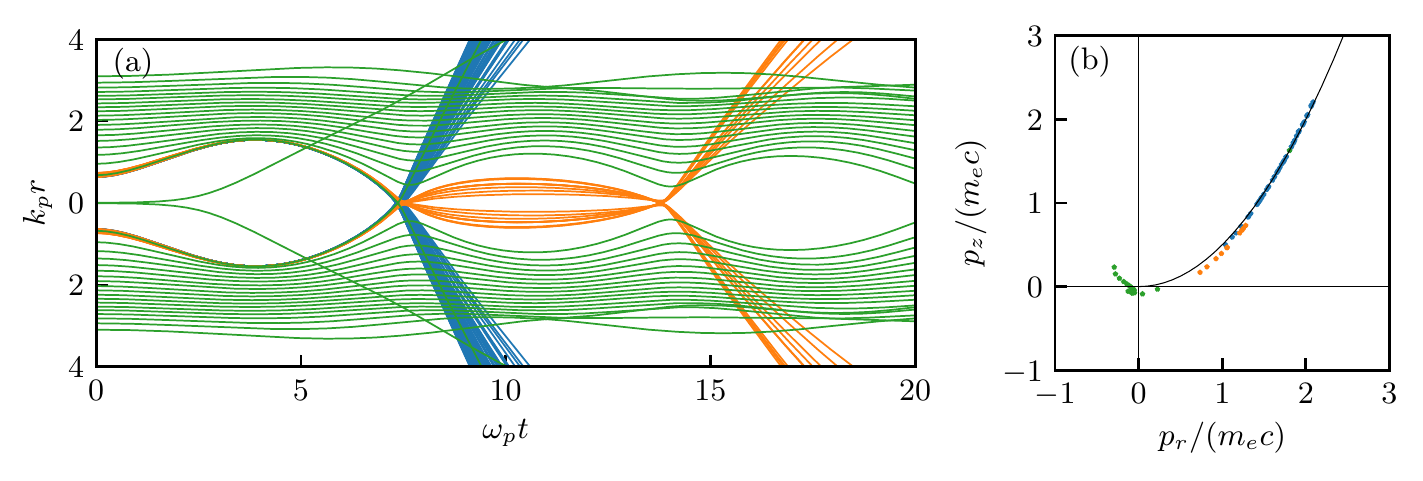}
	\caption{(a) Selected trajectories of plasma electrons in the co-moving window: blue (orange) lines correspond to electrons escaping after the first (second) bubble, and green lines to other electrons. (b) The position of these electrons at $\omega_p t = 18$ on the momentum plane ($p_r, p_z$). The thin line shows the parabola $p_z = p_r^2/(2m_e c)$.}\label{fig9-escape}
\end{figure*}

\section{Wave breaking}\label{sec4}

The wavebreaking in the considered strongly nonlinear regime leads to ejection of some electrons from the plasma with high (relativistic) velocities (figure~\ref{fig9-escape}). These electrons acquire a large momentum when the bubbles collapse and escape from the plasma column in the form of diverging tail waves. Other types of wavebreaking discussed in Ref.~\cite{PoP23-103112} do not result in such high energies. Most of the high-energy electrons appear in the tail of the first bubble. They carry away about 40\% of the wakefield energy and form a density ridge visible on the electron density map (figure~\ref{fig1-scheme}). Subsequent bubbles also generate tail waves, but weaker ones. The electrons ejected from different bubbles are initially located at approximately the same radii  (figure~\ref{fig9-escape}(a)). 

\begin{figure*}[tb]
	\centering
	\includegraphics[width = \textwidth]{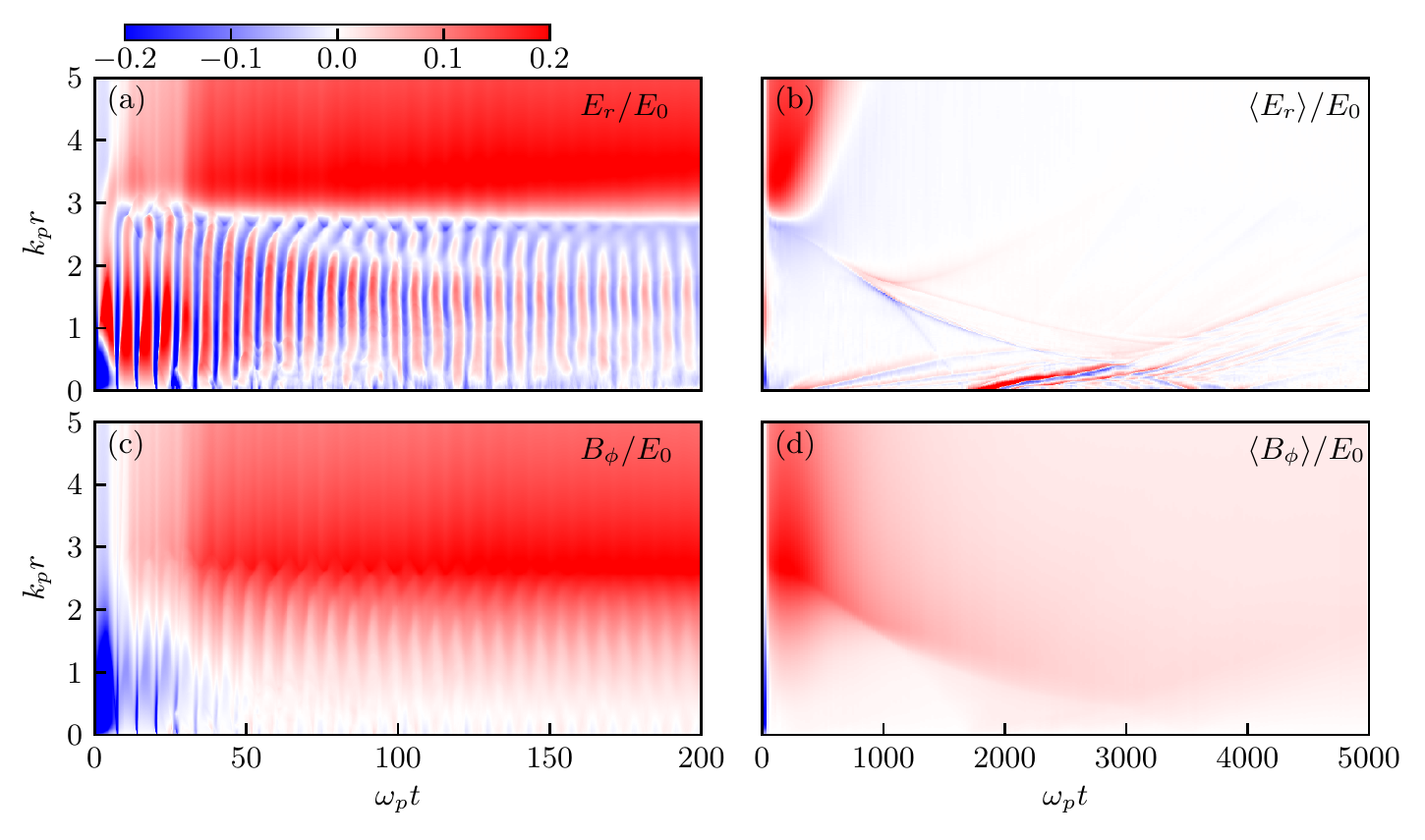}
	\caption{Evolution of (a),(b) radial electric field $E_r$ and (c),(d) azimuthal magnetic field $B_\phi$ at different timescales. At the longer timescale, the fields are time-averaged over $20\omega_p^{-1}$.}\label{fig10}
\end{figure*}

After the high-energy electrons leave the plasma, the excess positive charge concentrates in its outer layer (figure~\ref{fig3-total}) and generates a radial electric field around the plasma and inside this layer (figure~\ref{fig10}(a)). The layer thickness (of the order of $k_p^{-1}$) is much greater than the Debye length corresponding to the electron temperature \cite{JETP22-449}, because the near-boundary electrons oscillate (figures~\ref{fig1-scheme} and~\ref{fig3-total}) and respond to the average field as if their temperature were high. The ions located in this layer are then accelerated in the radial direction, forming the feature~3 in figure~\ref{fig2-big_dens}. As the plasma expands radially, the region of the strong electric field also moves to larger radii (figure~\ref{fig10}(b)). Similar processes occur when a moderately nonlinear plasma wave breaks \cite{PRL112-194801,PPCF63-055002} or the plasma is heated by a strong laser pulse \cite{RMP85-751}.

The escaping high-energy electrons also have a large positive longitudinal momentum, that is, they move in the direction of driver propagation. This is a common feature of wavebreaking, which follows from the basic wakefield properties. If the driver evolves slowly and propagates in an unperturbed plasma where the particles are initially in rest, then all plasma properties depend on the longitudinal coordinate $z$ and time $t$ only in their combination $\xi = z - ct$, and there is a relation between the relativistic factor $\gamma$ and the longitudinal momentum $p_z$ of plasma electrons \cite{PoP4-217}
\begin{equation}\label{e13}
    (\gamma - 1) m_e c^2 - e \Phi - c p_z = 0,
\end{equation}
where $\Phi$ is the so-called wakefield potential or pseudo-potential related to the longitudinal electric field $E_z$ as
\begin{equation}
    \Phi (r,\xi) = \int_\xi^\infty E_z (r, \xi') \, d\xi'.
\end{equation}
By introducing
\begin{equation}
    W = \frac{e \Phi}{m_e c^2} + 1
\end{equation}
we obtain from equation (\ref{e13})
\begin{equation}
    p_z = \frac{p_r^2}{2 m_e c W} + \frac{1 - W^2}{2W}.
\end{equation}
Outside the region of strong wakefields, $e\Phi \ll m_e c^2$, and $W \approx 1$. Therefore, any electron escaping from the plasma with a large radial momentum $p_r$ must also have a large positive longitudinal momentum
\begin{equation}
    p_z \approx \frac{p_r^2}{2 m_e c},
\end{equation}
and simulations confirm this (figure~\ref{fig9-escape}(b) and movie~2 in Supplementary materials).

\section{Plasma pinching}\label{sec5}

As forward-moving fast electrons leave the plasma column, an average current of the opposite direction appears inside the plasma. This current creates a strong azimuthal magnetic field (figure~\ref{fig10}(c),(d)), which in turn exerts an inward radial force on the return current of plasma electrons. A charge separation electric field (blue area at $\omega_p t \lesssim 500$ in figure~\ref{fig10}(b)) transmits this force to ions and accelerates them toward the axis. Thus, most of the ions move inward under the magnetic force, while a smaller (outer) part moves outward, entrained by the escaping fast electrons (figure~\ref{fig2-big_dens}(a)). The inward-directed force is nonlinear in radius and is strongest at the periphery, so the far ions overtake the near ones and form a compression wave (feature~4 in figure~\ref{fig2-big_dens}(a)), which behaves like a soliton and is similarly accompanied by charge-separation electric fields (figure~\ref{fig10}(b)). When high density spikes appear in the near-axis region, they generate ion-acoustic solitons by the mechanism discussed in section~\ref{sec3}. As a result, the dense plasma pinch formed near the axis is highly inhomogeneous (feature~5 in figure~\ref{fig2-big_dens}(a)). 

At even longer times, ionization of the surrounding neutrals occurs, and we observe an increase of ion density beyond the initial plasma radius (feature~6 in figure~\ref{fig2-big_dens}(a)). We will not discuss this feature, as it is described in detail in Ref.\,\cite{NatComm11-4753}.

\begin{figure}[tb]
	\centering
	\includegraphics[width=\linewidth]{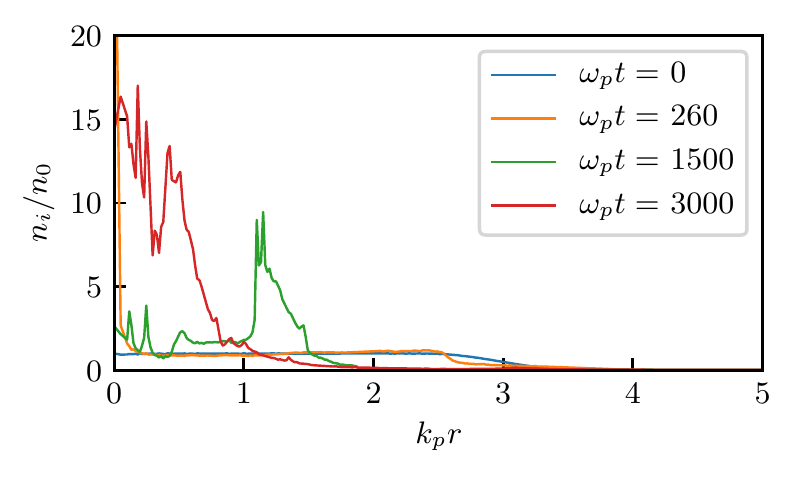}
	\caption{Radial profiles of the ion density at different times after the drive beam passage. }\label{fig11}
\end{figure}

\section{Discussion}\label{sec6}

The technique in which an additional beam or discharge passes through the plasma before the main pulse to form the desired density profile is widely used in plasma-based wakefield accelerators \cite{RMP81-1229}. The long-term evolution of a strongly nonlinear plasma wave can provide additional opportunities for this. We considered the case typical of a high-amplitude plasma wave in a finite-radius plasma and observed a variety of ion density profiles formed at different times after beam passage (figure~\ref{fig11}). These are a narrow density spike at the center of a nearly uniform plasma at $\omega_p t \sim 260$, a density channel with steep high-density walls at $\omega_p t \sim 1500$, and a narrow filament with an order of magnitude increased density at $\omega_p t \sim 3000$. A fine-scale density structure is superimposed on the latter two. At the heart of this variety are the four effects discussed. Two of them, the on-axis density peaking and the formation of ion-acoustic solitons, do not require a boundary and will also work in an unbounded plasma. Two others, the radial acceleration of sheath ions and the plasma pinching, are characteristic only of radially bounded plasmas.

\ack

The authors thank T.\,Silva and J.\,Vieira for providing the OSIRIS output data from \cite{NatComm11-4753}, which were used as input data for the simulations presented, and thank M.\,Hogan, A.\,Sosedkin, V.\,Yakimenko, and R.\,Zgadzaj for helpful discussions. This work was supported by the Russian Science Foundation, project 20-12-00062.

\end{document}



{\large Supplementary materials contain two movies}\\

{\bf Movie 1:}

The upper fragment (a) shows the radial momentum of plasma ions $p_r$ (orange points, left axis) and the radial electric field $E_r$ (blue line, right axis) versus radius $r$ at different times $t$.

The lower fragment (b) shows the densities of plasma ions $n_i$ (blue line) and electrons $n_e$ (orange line) versus radius $r$ at the corresponding times $t$.

{\bf Feature 1} (formation of the first density peak on the axis) is visible in (a) as the appearance of ions with $p_r < 0$ in the region $k_p r \lesssim 0.5$ at $\omega_p t \lesssim 200$ and in (b) as narrow peaks of $n_i$ and $n_e$ on the axis at $\omega_p t \sim 200$. The average electric field, which accelerates the ions, is not clearly visible against the field oscillations in the plasma wave.

{\bf Feature 2} (ion-acoustic solitons) are best seen in (b) as moving ion density peaks accompanied by less pronounced electron density peaks at $400 < \omega_p t < 1600$ in the region $k_p r \lesssim 0.5$. Fragment (a) shows that these peaks are accompanied by a radial electric field directed away from the peaks and by a positive perturbation to the ion radial momentum. Note that during the formation of each soliton, a group of ions drop out of the collective motion and freely propagates in the radial direction (the first event of this kind is seen as an isolated group of ions with $p_r \sim 15 m_e c$, which appears at $\omega_p t \approx 400$).

{\bf Feature 3} (ambipolar plasma expansion) is a consequence of the strong positive electric field visible in (a) at $k_p r \gtrsim 2.5$. Ions located in this region gain a large positive radial momentum, also visible in (a). The density of these ions is small and poorly seen in (b). However, we can notice in (b) that electron and ion density profiles differ at $\omega_p t \lesssim 300$, thus forming an uncompensated charge.

{\bf Feature 4} (compression wave) is seen in (b) as the ion and electron density peak moving toward the axis at $700 \lesssim \omega_p t \lesssim 3000$. Similarly to ion-acoustic solitons, the peak is accompanied by perturbations of the electric field.

{\bf Feature 5} (plasma pinching) occurs at $\omega_p t \sim 3000$ and is seen in (b) as a wide (up to $k_p r \sim 1$) peak of ion and electron densities. The peak has an internal structure that looks like a developed ion-acoustic turbulence, with each elementary wave creating its own perturbations of the electric field and ion velocity, seen in (a). After the maximum pinching stage, a collisionless damping of the ion waves leads to ion heating to about 2\,keV, which corresponds to the momentum range of $10m_e c$ in (a).

{\bf Feature 6} (ionization of neutrals) is seen in (a) as scattered orange points, a number of which increases with time. These points are newly appeared ions. Their density, however, is small on the considered timescales and almost invisible in (b).\\

{\bf Movie 2:}

Movie 2 is similar to figure 9 from the paper, but shows the temporal evolution of electron momenta.

The left fragment (a) shows selected trajectories of plasma electrons in the co-moving window. Blue (orange) lines correspond to electrons escaping after the first (second) bubble, and green lines to electrons remaining in the plasma during the considered time interval. 

The right fragment (b) shows the motion of these electrons on the momentum plane ($p_r, p_z$). The thin line shows the parabola $p_z = p_r^2/(2m_e c)$.